\documentclass{article}
\usepackage{jheppub}

\usepackage{color}
\usepackage{amsmath}

\usepackage{ulem}

\newcommand{\roughly}[1]{\mathrel{\raise.3ex\hbox{$#1$\kern-0.85em
\lower1ex\hbox{$\sim$}}}}

\newcommand{\lsim}{\roughly<}
\newcommand{\gsim}{\roughly>}

\def\exd{{\hbox{d}}}

\def\ba{\begin{eqnarray}}
\def\ea{\end{eqnarray}}
\def\be{\begin{equation}}
\def\ee{\end{equation}}

\def\bfr{{\bf r}}

\def\ssD{{\scriptscriptstyle D}}
\def\ssF{{\scriptscriptstyle F}}

\def\ssI{{\scriptscriptstyle I}}

\def\ssT{{\scriptscriptstyle T}}

\def\A{\mathcal{A}}
\def\B{\mathcal{B}}

\def\E{\mathcal{E}}

\def\I{\mathcal{I}}

\def\L{\mathcal{L}}
\def\O{\mathcal{O}}

\def\eff{{\rm eff}}

\def\nn{\nonumber}

\def\({\left(}
\def\){\right)}

\def\pref#1{(\ref{#1})}

\title{Power-counting during single-field slow-roll inflation}

\author[a]{Peter Adshead,}
\author[b,c]{C.P.~Burgess,}
\author[d]{R.~Holman,}
\author[e]{and Sarah Shandera}
\affiliation[a]{Department of Physics, University of Illinois at Urbana-Champaign, Urbana, IL 61801, USA }
\affiliation[b]{Physics \& Astronomy, McMaster University, Hamilton, ON, Canada, L8S 4M1}
\affiliation[c]{Perimeter Institute for Theoretical Physics, Waterloo, Ontario N2L 2Y5, Canada }
\affiliation[d]{Minerva Schools at KGI,1145 Market St, San Francisco, CA 94103, USA }
\affiliation[e]{Institute for Gravitation and the Cosmos, The Pennsylvania State University, University Park, PA 16802, USA}

\emailAdd{adshead@illinois.edu} 
\emailAdd{cburgess@perimeterinstitute.ca} 
\emailAdd{rh4a@andrew.cmu.edu} 
\emailAdd{shandera@gravity.psu.edu}

\date{}

\abstract {We elucidate the counting of the relevant small parameters in inflationary perturbation theory. Doing this allows for an explicit delineation of the domain of validity of the semi-classical approximation to gravity used in the calculation of inflationary correlation functions. We derive an expression for the dependence of correlation functions of inflationary perturbations on the slow-roll parameter $\epsilon = -\dot{H}/H^2$, as well as on $H\slash M_p$, where $H$ is the Hubble parameter during inflation. Our analysis is valid for single-field models in which the inflaton can traverse a Planck-sized range in field values and where all slow-roll parameters have approximately the same magnitude. As an application, we use our expression to seek the boundaries of the domain of validity of inflationary perturbation theory for regimes where this is potentially problematic: models with small speed of sound and models allowing eternal inflation.}

\begin{document}
\preprint{IGC-17/8-2}

\maketitle
\section{Introduction and summary}

There is considerable evidence from both the observed temperature fluctuations in the cosmic microwave background (CMB) \cite{CMB} and the distribution of large-scale structure \cite{LSS} across the observable universe that points to the existence of a specific pattern of near-scale-invariant primordial fluctuations. Furthermore, the properties of these primordial fluctuations inferred from observations seem well-described by quantum fluctuations \cite{Fluctuations} during an early accelerated epoch; usually assumed to be due to inflation \cite{firstINF}.  The success of this description argues that large-scale quantum-gravity effects are not only detectable but that they have, in fact, been detected: verily a triumph of modern physics.  

Any  quantum treatment of gravity eventually collides with its non-renormalizability \cite{NRGR}, which arises because its coupling constant (Newton's constant $8\pi G = 1/M_p^2$) has dimensions of inverse mass in fundamental units, for which $\hbar = c = 1$.  Because $G$ has negative mass-dimension, any perturbative series in $G$ inevitably becomes a low-energy expansion, since the dimensionless combination is $GE^2/2\pi = (E/4\pi M_p)^2$, for some $E \ll M_p$. Gravity (like other non-renormalizable theories \cite{EFTreview}) therefore naturally lends itself to a description in terms of an effective field theory (EFT) \cite{GREFT0, GREFT}. EFTs are relevant because they are designed specifically to  efficiently exploit the low-energy limit whenever a system enjoys a large hierarchy of scales, such as $E \ll M$ (with $M \sim M_p$). 

The lagrangian of an EFT is usually a complicated expression and, when written order-by-order in powers of the heavier mass, is usually an arbitrary real, local function of all possible combinations of the fields and their derivatives consistent with the symmetries of the problem. It is because powers of derivatives come together with inverse powers of the heavy scale, $M$  --- where usually (though, as we see below, not exclusively) for gravity $M \sim M_p$ --- that this complicated lagrangian is useful. Only a few interactions prove to be relevant at low orders in the $1/M$ expansion. 

Indeed, the key question one asks with {\it any} EFT is: precisely which interactions appear in exactly which ways inside which graphs within a graphical expansion at any fixed order in the low-energy ratio $E/M$? Answering this question is called `power counting' the EFT, and the answer is important for two reasons. First, (as applied to the contributions you compute) it is central to being able to systematically exploit the predictions of the EFT because it identifies precisely which interactions enter to any given desired order in the low-energy expansion. Second, (as applied to the contributions you do {\it not} compute) it shows precisely what is the theoretical error by showing how large are the contributions due to the leading terms being neglected. Power-counting arguments quantify the size of the theoretical error, and this is clearly a prerequisite for any meaningful comparison with observations.

But, as we review below, for gravity there is also a bonus: it is this kind of calculation that shows why semi-classical calculations are usually (but not always) a good approximation in gravitating systems.  For example, it reveals the semi-classical expansion during cosmology to be a series in powers of $(H/4\pi M_p)^2$, where $H$ is the spacetime's Hubble parameter. This answers a question that is probably not asked often enough: why is a classical analysis valid in the first place, and where does it start to break down?

Power-counting is also related to, but logically distinct from, issues of technical naturalness.\footnote{See {\it e.g.} ref.\ \cite{CCRev} for a review --- similar in spirit to the approach used here --- of what technical naturalness is and why it is regarded as a useful criterion when developing theories describing Nature.} It is related inasmuch as technical naturalness asks whether the sizes of effective couplings change as one integrates out states of various mass, with the particular focus being integrating out the most energetic states since these are the ones that potentially contribute dangerously to a few vulnerable low-energy quantities (such as small scalar masses or small vacuum energies). Power counting is relevant to addressing naturalness questions because the point of a power-counting argument is to identify systematically how various large mass scales contribute to any particular observable. 

But power counting and technical naturalness are distinct because although it can be an uncomfortable embarrassment (that one usually feels requires explaining) to have a theory that is not technically natural, the theory itself remains a self-consistent tool in which one can make sensible theoretical predictions. By contrast, if power counting indicates an uncontrolled dependence on a large mass then this really signals our inability to make quantitative predictions at all (at least purely within the low-energy limit). If a theory is in a regime where power-counting does not allow an expansion in small energy ratios it might or might not be technically natural; we simply do not know since we cannot compute with it reliably enough to tell. But if a theory is not technically natural its renormalized parameters are being chosen in an odd way that begs for an explanation, but this in itself need not undermine our understanding of how to make predictions (or, as is sometimes argued, of EFT methods in general).

Since inflationary predictions of primordial fluctuations might provide the first observation of a quantum gravity effect, power-counting during inflation is particularly important. In this note we provide the power-counting expression for the simplest single-field slow-roll models in which all slow-roll parameters are similar in size. We start, in \S\ref{sec:PCgrav}, by using a brief review of standard results for semiclassical power-counting in pure gravity to define notation useful later. (EFT aficionados should feel free to skip this section.) For inflationary applications there are a variety of things called EFTs \cite{IEFTWbg,InfEFT,InfEFT2} and our power-counting arguments apply to most of them. Although we explicitly make our arguments for a scalar-metric theory expanded about a rolling background --- closer in spirit to ref.\ \cite{IEFTWbg} --- we believe our arguments also go through for the more general framework of fluctuations about single-clock backgrounds given in ref.\ \cite{InfEFT}.

Our main inflationary result is derived in \S\ref{sec:PCinf}, with \S\ref{ssec:semiclass} tracking the low-energy factors that control the semiclassical approximation and \S\ref{ssec:slowroll} adding the additional information about dependence on slow-roll parameters. These sections culminate in eq.~\pref{PCinf}, which expresses how any graph contributing to a connected $n$-point correlation function (at horizon exit, $k = aH$) depends on the two small parameters: the slow-roll parameters, $\epsilon$, and the ratio of $H$ to other, higher, scales. Besides reproducing standard results --- as is argued is true for the lowest few $n$-point functions in \S\ref{ssec:examples} --- this expression allows the determination of when the underlying semiclassical expansion breaks down, thereby allowing a systematic inference of the boundaries of its domain of validity. Simpler `unitarity arguments' can also identify which small parameter controls perturbation theory; however, a full power-counting result does more because it systematically identifies how many powers of each small parameter enters from any particular graph in a perturbative expansion.   

We use the power-counting formula, eq.\ \pref{PCinf}, to explore the edges of the perturbative regime in two different ways. First, in \S\ref{ssec:smallcs}, we explore `small-$c_s$' models, in which the propagation of signals can occur with speeds much smaller than the speed of light. We show why these theories push the envelope of the vanilla power-counting arguments used here, and sketch how these arguments might be extended to include a small-$c_s$ regime in a controlled way, particularly for models like DBI inflation \cite{DBI} that enjoy additional symmetries.\footnote{We also in this section briefly discuss some of the other dangers that small $c_s$ models must address in specific realizations.} 

Finally, \S\ref{ssec:eternal} discusses the regime of eternal inflation, which we argue also probes calculational boundaries in interesting ways. In this case we argue that although semiclassical perturbation theory usually is in very good shape, for sufficiently small slow-roll parameters the $\epsilon$-expansion can become subdominant to the semiclassical expansion. In this regime it can be inconsistent to include $\epsilon$-dependence without also including quantum effects and higher-derivative corrections to General Relativity. 

\section{Power-counting with gravity}
\label{sec:PCgrav}

Power counting is the central step when working with EFTs, since it systematically identifies which of the many effective interactions can contribute to observables at any order in the small quantities underlying the EFT expansion  (for reviews, see for instance ref.\ \cite{EFTreview}). This section briefly summarizes standard results when these methods are applied to gravity \cite{GREFT0, GREFT}, before generalizing to include the effects of small slow-roll parameters. The reader familiar with this story should feel free to skip ahead to \S\ref{sec:PCinf}.

\subsection{GREFT}

General relativity is not renormalizable. Although this was regarded as a problem back in the day, we now know that non-renormalizability in itself is not that remarkable since other very predictive theories (like the Fermi theory of weak interactions or low-energy interactions of pions) also share this property. Non-renormalizability is generic whenever there are couplings (like Newton's constant, $G$, or the Fermi constant $G_\ssF$) with engineering dimensions that are inverse powers of mass, and the central observation that gives them predictive power is that any series in this coupling is necessarily a low-energy expansion: one works in powers of $GE^2$ where $E$ is a typical scale in the observables of interest. (The discussion below follows closely the treatment in ref.\ \cite{GREFT}.)

EFTs are the natural language for describing this sort of low-energy expansion, since they organize interactions from the get-go into a derivative expansion in order to identify most efficiently those that dominate at low energies. Given that gravity is described by the spacetime metric, $g_{\mu\nu}$, the effective lagrangian describing pure gravity to which one is led in this way is
\ba \label{GReffdef}
 - \frac{ \L_\eff}{\sqrt{-g}} &=& \lambda + \frac{M_p^2}{2} \, g^{\mu\nu} \, R_{\mu\nu} \nn\\
 && \qquad + \Bigl[  a_{41} \, R^2 + a_{42} \, R_{\mu\nu} R^{\mu\nu} + a_{43} \, R_{\mu\nu\lambda\rho} R^{\mu\nu\lambda\rho} +  \cdots \Bigr] \nn\\
 && \qquad\qquad\qquad +   \frac{1}{M^2}  \Bigl[  a_{61} \, R^3 + a_{62} \, R R_{\mu\nu} R^{\mu\nu} + \cdots \Bigr] + \cdots
\ea
corresponding to a sum over all possible curvature invariants. In this expression the first line represents the usual terms of General relativity --- consisting of a cosmological constant, $\lambda$, and the Einstein-Hilbert action --- while the second line contains curvature-squared terms, the third line curvature-cubed terms, and so on. We have introduced the reduced Planck mass, $M_p^2=(8\pi G)^{-1}$.

As written, the couplings $a_{di}$ are dimensionless (in 4 dimensions) and to this end an appropriate power of an overall mass scale $M$ is factored out of the curvature-cubed and higher terms. Two things are important in this expansion: the dimensionless quantities $a_{di}$ are usually at most order-unity;\footnote{This property is {\it not} true for some popular theories, like Higgs inflation \cite{Nonminscalar,HI} or curvature-squared inflation \cite{Staro}, and it is this property that makes these theories difficult to obtain from known UV completions \cite{InfPowerCount, HIdiff, UVblock}.} and the scale $M$ is likely to be much smaller than $M_p$. To see why these are true, imagine obtaining these interactions by integrating out a heavy particle of mass $M$. Within perturbation theory such a calculation generically predicts $M$ appears as required on dimensional grounds, and that the $a_{di}$ are proportional to dimensionless couplings and suppressed by powers of $2\pi$ (see below for more details on why). 

One also might expect that integrating out a particle of mass $M$ would contribute an amount $\delta M_p^2 \propto M^2$ to the Einstein-Hilbert action, as well as an amount $\delta \lambda \propto M^4$ to the cosmological constant. But, the correction to the Einstein-Hilbert term is negligible if $M \ll M_p$ (unlike the contribution $R^3/M^2$, which completely swamps any prior contribution of order $R^3/M_p^2$). It is an unsolved puzzle why the cosmological constant is not also dominated by contributions from the largest values of $M$, so we simply drop $\lambda$ until discussing inflationary models in later sections.

Finally, it should also be noted that many of the interaction terms in the action in eq.\ \pref{GReffdef} are redundant, in that their coefficients do not appear independently in observables. Two common reasons for this are if the term in question is a total derivative (such as a $\Box R$ term, not written explicitly above) or if the term can be removed by performing a field redefinition. As argued in more detail in ref.\ \cite{GREFT}, in practice this latter criterion means that terms can be dropped that vanish when using the lowest-order field equations (such as the vacuum Einstein equations in the present case). For pure gravity with $\lambda = 0$ this allows the dropping of any terms involving the undifferentiated Ricci tensor or Ricci scalar ($R_{\mu\nu}$ or $R = g^{\mu\nu}R_{\mu\nu}$). 

\subsection{Semiclassical perturbation theory}

For the purposes of estimating how these interactions contribute to observables, we imagine working in semiclassical perturbation theory. This involves expanding about a classical solution,
\be
 g_{\mu\nu} (x) = \hat g_{\mu\nu} (x) +
 \frac{h_{\mu\nu}(x)}{M_p} \,,
\ee
and rewriting \pref{GReffdef} as a sum of effective interactions
\be \label{GReffphih}
 \L_\eff = \hat \L_\eff + M^2 M_p^2 \sum_{n}
 \frac{c_{n}}{M^{d_{n}}} \; \O_{n} \left(
 \frac{ h_{\mu\nu}}{M_p} \right) \,,
\ee
where $\hat\L_\eff = \L_\eff(\hat g_{\mu\nu})$ is the lagrangian density evaluated at the background configuration.

The sum over $n$ runs over the labels for a complete set of interactions, $\O_{n}$, each of which involves $N_n \ge 2$ powers of the field $h_{\mu\nu}$.  ($N_n \ne 1$ because of the background field equations satisfied by $\hat g_{\mu\nu}$.) The parameter $d_{n}$ counts the total number of derivatives appearing in $\O_n$ (acting either on the background or the perturbation), and so the factor $M^{-d_n}$ is what is required to keep the
coefficients, $c_n$, dimensionless.  For instance, an interaction like 
\be
  \frac{c_n}{M_p} \; h^{\mu\nu} \hat\nabla_\lambda h_{\nu\rho} \hat\nabla^\lambda {h^\rho}_\mu, 
\ee
with indices raised and covariant derivatives built using the background metric, $\hat g_{\mu\nu}$, would have $d_n = 2$ and $N_n = 3$.   The overall prefactor, $M^2 M_p^2$, is chosen so that the kinetic terms --- {\it i.e.} those terms in the sum for which $d_n = N_n = 2$ --- are $M$ and $M_p$ independent. As is clear from the example, the operators $\O_n$ depend implicitly on the classical background, $\hat g_{\mu\nu}$, about which the expansion is performed.

The coefficients $c_n$ are calculable in terms of the $a_{di}$, but if $M \ll M_p$ the $c_n$'s cannot all be order unity. Comparing eqs.~\pref{GReffdef} and \pref{GReffphih} shows that the absence of $M_p$ in all of the curvature-squared and higher terms in \pref{GReffdef} implies the $c_n$ for these interactions should be of order
\be \label{cndgt2h}
 c_n = \left( \frac{ M^2}{ M_p^2} \right) g_n
 \qquad \hbox{(if $d_n > 2$)} \,,
\ee
where $g_n$ is at most order-unity and independent (up to logarithms) of $M$ and $M_p$. 

Perturbation theory proceeds by separating $\L_\eff - \hat \L_\eff$ into quadratic and higher order parts,
\be
 \L_\eff = \Bigl( \hat \L_\eff + \L_0 \Bigr)
 + \L_{\rm int} \,,
\ee
where $\L_0$ consists of those terms in $\L_\eff$ for which $N_n = 2$ and $d_n \le 2$. All other terms are lumped into $\L_{\rm int}$. Expanding the path integral in powers of $\L_{\rm int}$ allows the integral over $h_{\mu\nu}$ to be expressed as a sum of Gaussian integrals, classifiable in terms of Feynman graphs, with $\L_0$ defining the propagators of these graphs and $\L_{\rm int}$ their vertices in the usual way. Standard arguments show that this is a semiclassical expansion inasmuch as each loop corresponds to an additional order in $\hbar$ (though tracking powers of $\hbar$ in this way does not in itself yet identify the semiclassical parameter whose smallness makes the loop expansion a good approximation). 

To identify more cleanly what parameter controls the loop expansion we make the following dimensional argument. Imagine computing an {\it amputated} Feynman graph, whose $\E$ external lines are removed and so carry no dimensions. The propagators, $G(x,y)$, associated with each of the $\I$ internal lines in this graph come from inverting the differential operator appearing in $\L_0$. What matters about these for the present purposes is that they do not depend on $M$ and $M_p$, although they can depend on scales (like the Hubble scale, $H$) that arise in the background configuration, $\hat g_{\mu\nu}$.

The factors of $M$ and $M_p$ all come from vertices in the Feynman graph of interest, since they all come from terms in $\L_{\rm
int}$. Each time the interaction $\O_n$ contributes a vertex to the graph it comes with a factor of $c_n M_p^{2 - N_n} M^{2-d_n}$. If the graph contains $V_n$ number of vertices of type $n$ it therefore acquires a factor
\be
 \prod_n \Bigl[ c_n M_p^{2 - N_n} M^{2 - d_n} \Bigr]^{V_n}
 = M_p^{2 - 2L - \E} \, \prod_n \Bigl[  c_n M^{2 - d_n}
 \Bigr]^{V_n}  \,,
\ee
where the equality uses the identity
\be \label{endcons}
 2\I + \E = \sum_n N_n V_n
\ee
that expresses that the end of each line in the graph must occur at a vertex, as well as the definition,
\be \label{loopdef}
 L =1 + \I - \sum_n V_n  \,,
\ee
of the number of loops, $L$, of the graph.

Given this identification of how $M$ and $M_p$ arise, the dependence of the Feynman graph of interest on the other, low-energy, scales defined by the graph's external lines is set by evaluating the Feynman rules. The result is particularly simple in the special case where there is only one such a low-energy scale, since in this case its appearance is dominantly determined (up to logarithms) on dimensional grounds. Because dimensional arguments are complicated by ultraviolet divergences --- arising due to the singularities in the propagators, $G(x,y)$, in the coincidence limit $y \to x$ ---for the purposes of making the dimensional argument it is very convenient to regularize these divergences using dimensional regularization. In this case all divergences arise as poles as the spacetime dimension approaches 4, and the overall dimension of the graph is set by the physical scale appearing in the external legs (such as the scale $H$ characterizing the size of a derivative of the background classical configuration).

Denoting this external scale by $H$, the contribution of a graph involving $\E$ (amputated) external lines, $L$ loops and $V_n$ vertices involving $d_n$ derivatives becomes
\be \label{PCresult0}
 \A_\E (H) \simeq H^2 M_p^2 \left( \frac{1}{M_p} \right)^\E
 \left( \frac{H}{4 \pi \, M_p}
 \right)^{2L} \prod_n \left[ c_n \left( \frac{H}{M}
 \right)^{d_n-2} \right]^{V_n} \,,
\ee
with factors of $4\pi$ also included using standard arguments (see, {\it e.g.}~\cite{GREFT}).  Keeping in mind the factors of $M$ and $M_p$ hidden in some of the $c_n$'s --- {\it c.f.} eq.~\pref{cndgt2h} --- it is useful to separate out the interactions with more than two derivatives in this result, to get
\be \label{PCresultGR}
 \A_\E (H) \simeq H^2 M_p^2 \left( \frac{1}{M_p} \right)^\E
 \left( \frac{H}{4 \pi \, M_p}
 \right)^{2L} \left[ \prod_{d_n = 2} c_n^{V_n} \right]
  \prod_{d_n \ge 4} \left[ g_n \left( \frac{H}{M_p}
 \right)^2 \left( \frac{H}{M}
 \right)^{d_n-4} \right]^{V_n}  \,.
\ee
Notice the dimension of $\A_\E$ here is what would be expected for the coefficient of $\phi^\E$ in an expansion of the one-particle irreducible (1PI) action: {\it i.e.}~$\A_\E$ has dimension (mass)${}^{4-\E}$, as appropriate given its external lines have been amputated.

Eq.~\pref{PCresultGR} identifies which combination of scales justifies regarding quantum effects to be small enough to allow semi-classical methods in General Relativity. A generic necessary condition for graphs with more loops (for a fixed number of external lines) to be parametrically suppressed compared to those with fewer loops is to have $H$ be small enough to ensure
\be \label{Loopcond}
 \frac{H}{4 \pi M_p} \ll 1 \,,
\ee
while the suppression of interactions coming from higher-derivative interactions additionally requires
\be
 g_n \left( \frac{H}{M_p}
 \right)^2 \left( \frac{H}{M}
 \right)^{d_n-4} \ll 1 \quad \hbox{(for $d_n \ge 4$)}\,.
\ee
Repeated insertions of two-derivative interactions ({\it i.e.}~those coming from the Einstein-Hilbert action) do not generically generate large contributions because these satisfy
\be
 c_n \simeq 1 \quad \hbox{(for $d_n = 2$)} \,.
\ee
The lack of suppression of these interactions expresses the principle of equivalence, since it shows that they are all generically equally important in a given low-energy process.

Eq.~\pref{PCresultGR} is central to {\it all} applications of General Relativity, since it identifies systematically which interactions are important in any given physical process. In particular, because $H$ is always in the numerator it shows that control over semiclassical methods always requires a low-energy approximation (as is indeed expected due to the presence of the dimensionful non-renormalizable coupling, $G$). 

In practice, most people using General Relativity treat it as a classical field theory and eq.\ \pref{PCresultGR} shows why this is usually valid: for a fixed number of external lines the dominant contributions come if $L=0$ and $V_n = 0$ for all interactions for which $d_n > 2$. Since $d_n$ is even (and the assumption that $\lambda$ is negligible implies $d_n \ge 2$), this means that the dominant processes are those computed at tree level using only interactions with precisely $d_n=2$ derivatives; that is, classical processes computed using only the Einstein-Hilbert action. 

But eq.\ \pref{PCresultGR} also shows which interactions contribute at subleading order: the dominant corrections are those using only $d_n = 2$ interactions but with $L=1$, or those with $L=0$ for which the $d_n=2$ interactions are supplemented with only a single $d_n=4$ interaction. That is, the next-to-leading contribution is suppressed compared to classical General Relativity by $(H/4 \pi M_p)^2$ and comes from one-loop General Relativity plus tree graphs containing exactly one curvature-squared interaction. It is the coefficient of the tree-level, curvature-squared interactions that renormalize the UV divergences that arise in the one-loop graphs. 

And so on, to any required accuracy in powers of $H/M$ and $H/M_p$.

\section{Power-counting in simple inflationary models}
\label{sec:PCinf}

We now repeat and extend the above arguments to simple inflationary models, following the treatment in ref.\ \cite{InfPowerCount}. 

\subsection{Scalar-metric models}

We start by adding $N$ dimensionless scalar fields, $\theta^i$, though later restrict to single-field models. The effective lagrangian obtained as a derivative expansion is then
\ba \label{Leffdef}
 - \frac{ \L_\eff}{\sqrt{-g}} &=& v^4 V(\theta) + \frac{M_p^2}{2}
 \, g^{\mu\nu} \Bigl[  W(\theta) \, R_{\mu\nu}
 + G_{ij}(\theta) \, \partial_\mu \theta^i
 \partial_\nu \theta^j \Bigr] \\
 && \quad + A(\theta) (\partial \theta)^4 + B(\theta)
 \, R^2 + C(\theta) \, R \, (\partial \theta)^2
 + \frac{E(\theta)}{M^2} \, (\partial \theta)^6
 + \frac{F(\theta)}{M^2} \, R^3 + \cdots \,,\nn
\ea
with terms involving up to two derivatives written explicitly and the rest written schematically, inasmuch as $R^3$ collectively represents all possible independent curvature invariants involving six derivatives, and so on. As in the previous section, the explicit mass scales $v$, $M_p$ and $M$ are extracted so that the functions $V(\theta)$, $W(\theta)$, $A(\theta)$, $B(\theta)$ {\it etc}, are dimensionless. Eq.~\pref{Leffdef} normalizes the scalar fields so that their kinetic term has Planck mass coefficient. With inflationary applications in mind we take $M \ll M_p$, and for inflationary applications we take $V \simeq v^4 \ll M^4$ when $\theta \simeq \O(1)$. 

Again expanding about a classical solution,
\be
 \theta^i(x) = \vartheta^i(x) + \frac{\phi^i(x)}{M_p}
 \quad \hbox{and} \quad
 g_{\mu\nu} (x) = \hat g_{\mu\nu} (x) +
 \frac{h_{\mu\nu}(x)}{M_p} \,,
\ee
allows the lagrangian in eq.\ \pref{Leffdef} to be written as 
\be \label{Leffphih}
 \L_\eff = \hat \L_\eff + M^2 M_p^2 \sum_{n}
 \frac{c_{n}}{M^{d_{n}}} \; \O_{n} \left(
 \frac{\phi}{M_p} , \frac{ h_{\mu\nu}}{M_p} \right)
\ee
where as before $\hat\L_\eff = \L_\eff(\vartheta,\hat g_{\mu\nu})$ and the interactions, $\O_{n}$, involve $N_n = N^{(\phi)}_n + N^{(h)}_n \ge 2$ powers of the fields $\phi^i$ and $h_{\mu\nu}$. Also as before the parameter $d_{n}$ counts the number of derivatives appearing in $\O_n$, the coefficients $c_n$ are dimensionless and the prefactor, $M^2 M_p^2$, ensures the kinetic terms (and so also the propagators) are $M$ and $M_p$ independent. 

Following the steps of the previous section, we assign $M$ and $v$ dependence to the coefficients $c_n$ (for $d_n \ne 2$) so that eq.\ \pref{Leffphih} captures the same dependence as does eq.\ \pref{Leffdef}. For simplicity, we first treat derivatives of the background and fluctuations on equally footing, as was done in the previous section. We relax this assumption in the next section. For $d_n > 2$ this implies $c_n$ is given by eq.\ \pref{cndgt2h} where $g_n$ is order unity, and for terms with no derivatives --- {\it i.e.}~those coming from the scalar potential, $V(\theta)$ --- we have
\be \label{cndeq0}
 c_n = \left( \frac{v^4}{M^2 M_p^2} \right) \lambda_n
 \qquad \hbox{(if $d_n = 0$)} \,,
\ee
where the dimensionless couplings $\lambda_n$ are independent of $M_p$ and $M$ (and are related to slow-roll parameters in what follows). 

In terms of the $\lambda_n$'s the above assumptions imply the scalar potential has the schematic form\footnote{Notice our assumption that $\phi$ is normalized by $M_p$ implies a qualitative steepness for the scalar potential: $V = v^4 U(\phi/M_p)$ where $U(x)$ is order unity when evaluated at order-unity arguments. Technical naturalness asks whether this form remains valid after integrating out other states (see  ref.\ \cite{InfPowerCount} for how the power-counting arguments used here can help assess this).} 
\be
 V(\phi) = v^4 \left[ \lambda_0 + \lambda_2 \left(
 \frac{\phi}{M_p} \right)^2 + \lambda_4 \left(
 \frac{\phi}{M_p} \right)^4 + \cdots \right] \,,
\ee
which shows that we choose $V$ to range through values of order $v^4$ as $\phi^i$ range through values of order $M_p$. When applied to inflationary models in later sections we make the further `slow-roll' assumption that constrain $\lambda_n$ to be smaller than order unity.   Although these choices do not capture all possible inflationary models, they do capture those for which the inflaton rolls over a Planckian range and for which all slow-roll parameters are of similar size, such as single-field models with observable tensor-to-scalar ratio, $r$. 

The natural scale for the scalar masses under the above assumptions is $m \simeq \sqrt{V''}/M_p \simeq \sqrt\epsilon \; v^2/M_p \simeq \sqrt\epsilon \; H$, where the natural value for the Hubble scale is $H \simeq \sqrt{V}/M_p \simeq v^2/M_p$. Assuming the classical background is described by slow-roll inflation the derivatives of the canonically normalized scalar fields $\varphi^i = M_p \vartheta^i$ also satisfy
\be \label{SReqn}
 \dot\varphi \simeq \frac{V'}{H} \simeq \frac{M_p V'}{\sqrt{V}}
 \simeq \sqrt{\epsilon V} \simeq \sqrt\epsilon \; v^2 \simeq \sqrt\epsilon \; H M_p\,.
\ee

\subsection{Semiclassical perturbation theory}
\label{ssec:semiclass}

Our goal is to identify how any observable depends on both the small energy ratios $H/M$ and $H/M_p$ as well as the slow-roll parameter, $\epsilon$. To start off let us take $\epsilon \simeq \O(1)$ and purely count powers of $H/M_p$, by repeating the dimensional power-counting argument of earlier sections. In subsequent sections we dial down $\epsilon$ to examine its competition with $H/M_p$.

To this end we expand $\L_\eff = \Bigl( \hat \L_\eff + \L_0 \Bigr) + \L_{\rm int}$, and examine the size of a Feynman graph having $\E$ external lines, with external lines characterized by the single low-energy scale $H$. We track the powers of $M$, $v$ and $M_p$ coming from the vertices and determine the $H$-dependence on dimensional grounds, in the manner that led to \pref{PCresult0}, including making explicit the factors of $v$, $M$ and $M_p$ hidden in the $c_n$'s when $d_n \ne 2$. 

For cosmological applications it also proves useful to normalize amplitudes differently than in \pref{PCresult0}. Whereas the previous section normalizes $\A_\E$ as appropriate to amputated Feynman graphs, for cosmology it is more useful to track correlation functions\footnote{For cosmology one usually also separately tracks dependence on $H$ and mode momentum $k/a$, but these are the same size if we assume all momentum components have a similar size and are evaluated during the epoch of most interest: horizon exit.} for which we attach a propagator to each external line and integrate over the space-time location of the amputated graph. Since power-counting here associates a factor of $H$ for each dimension, the resulting Feynman amplitude scales with the parameters according to $\B_\E \simeq \A_\E H^{2\E-4}$.

Combining everything leads to the result
\ba \label{PCresult}
 \B_\E (H) &\simeq& \frac{M_p^2}{H^2} \left( \frac{H^2}{M_p} \right)^\E
 \left( \frac{H}{4 \pi \, M_p}
 \right)^{2L} \left[ \prod_{d_n = 2}  c_n^{V_n} \right] \\
 && \qquad \qquad \qquad \times
 \left[ \prod_{d_n = 0}  \lambda_n^{V_n} \right]
 \prod_{d_n \ge 4} \left[ g_n \left( \frac{H}{M_p}
 \right)^2 \left( \frac{H}{M}
 \right)^{d_n-4} \right]^{V_n}  \,, \nn
\ea
which uses $H \simeq v^2/M_p$ to rewrite the potentially dangerous $d_n = 0$ term as
\be
 \prod_{d_n = 0} \left[ \lambda_n \left( \frac{v^4}{H^2 M_p^2}
 \right) \right]^{V_n} \simeq \prod_{d_n=0} \lambda_n^{V_n} \,.
\ee
Although insertions of scalar interactions can sometimes undermine the underlying $H/M_p$ expansion \cite{InfPowerCount}, this does not happen for potentials of the form assumed in inflationary models.

Eq.~\pref{PCresult} shows that, under the assumptions given, the presence of scalar fields does not undermine the validity of the underlying semiclassical expansion, which again relies on the low-energy approximation \pref{Loopcond}: $H \ll 4 \pi M_p$. Just as for pure gravity the leading contribution comes from classical physics, though this time using the zero- and two-derivative parts of the action. This is what justifies standard classical treatments of inflation. Again as before the dominant subleading terms are down by $(H/4\pi M_p)^2$ \cite{EFTadiabatic1} and arise at one-loop together with classical contributions with appropriate counterterms with up to four derivatives. 

These power-counting results can be used to study the sensitivity of the inflationary choices made when specifying the effective action, with the generic result that integrating out a heavy field of mass $m$ gives contributions of the same form as those already found in the action, but with $v \simeq M \simeq m$ \cite{InfPowerCount, EFTadiabatic}, together with potential non-adiabatic corrections that can invalidate the underlying EFT description \cite{EFTnonadiabatic, GREFT}.

\subsection{Slow-roll suppression}
\label{ssec:slowroll}

Generically, derivatives of the background and derivatives of the fluctuations may be of parametrically different sizes. In inflation, this occurs because derivatives of background fields are additionally suppressed by powers of the slow-roll parameters. We now dial down the generic slow-roll parameter, $\epsilon$, and find that there are two ways that $\epsilon$ modifies eq.\ \pref{PCresult}. 

First, the assumed flatness of the inflationary potential (assuming all slow-roll parameters to be of the same order of magnitude) allows us to write the $s$th derivative of the scalar potential as $(\exd^sV/\exd \varphi^s) \simeq \epsilon^{s/2} \, V/M_p^s \simeq \epsilon^{s/2} \; v^4/M_p^s$ and so 
\be
   \lambda_n \simeq \epsilon^{N_n/2} \hat \lambda_n \,,
\ee
where $\hat \lambda_n$ is order unity and $N_n$ counts the number of lines that meet at the vertex in question.  Using this in eq.\  \pref{PCresult} shows how insertions of scalar interactions do not change the powers of $H/M_p$ but always cost powers of the slow-roll parameters,\footnote{Of course this conclusion relies on the assumed form $V = v^4 U(\phi/M_p)$ where $U(x)$ is order unity when evaluated at order-unity arguments.} with a factor of $\sqrt\epsilon$ arising for each scalar line that meets in the vertex.

The other way slow-roll parameters enter into eq.\ \pref{PCresult} is through scalar background-field derivatives, which we assume satisfy eq.\ \pref{SReqn} and its higher slow-roll extensions 
\be
  \frac{\exd^n \varphi}{ \exd t^n} \simeq \epsilon^{n/2} H^n M_p \,.
\ee
This kind of suppression arises once scalar-field derivatives are expanded about their background (assuming all slow-roll parameters are similar in size), as in $\partial_\mu( \varphi + \phi) = \dot \varphi \, \delta^0_\mu + \partial_\mu \phi$ and so on. (We assume negligible background gradient energy in $\varphi$ as required for inflation.)

To track these factors we replace the two labels $(d,i)$ counting the numbers of derivatives and fields in an interaction with five labels that do so separately for background and fluctuating fields: $(d,i;  D, I_s, I_h)$. Here $i$ and $d$ respectively denote the number of powers of background fields $\varphi$ (but not $\hat g_{\mu\nu}$) appearing in the vertex and the number of times these background fields are differentiated. The quantity $I_s$ similarly counts the number of powers of the fluctuating scalar fields ($\phi$) and $I_h$ counts the number of powers of metric fluctuations $h_{\mu\nu}$), while $D$ counts the total number of derivatives except those that act on the background scalar field (and so are separately counted by $d$). We incorporate the slow-roll suppression due to background evolution by requiring any vertex with these labels to be suppressed by 
\ba
 c_n &\to& c^{d,i}_{\ssD,\ssI_s,\ssI_h} \simeq \epsilon^{d/2} \hat c^{d,i}_{\ssD,\ssI_s,\ssI_h} \quad \hbox{(for $d_n = d+D = 2$)} 
 \nn\\
  \hbox{and} \qquad
 g_n &\to& g^{d,i}_{\ssD,\ssI_s,\ssI_h} \simeq \epsilon^{d/2} \hat g^{d,i}_{\ssD,\ssI_s,\ssI_h} \quad \hbox{(for $d_n = d+D > 2$)}   \,,
\ea
where the $\hat c$ and $\hat g$ are order-unity constants. Notice we assume no slow-roll suppression when the scalar field appears undifferentiated in the action anywhere except for the scalar potential, so that (for example) there is no additional suppression by powers of $\epsilon$ associated with any $\vartheta$-dependence in\footnote{Ignoring $\epsilon$ suppression in $W(\vartheta)$ likely over-estimates its size in models where the small size of $V(\vartheta)$ is understood because $\vartheta$ is a pseudo-Goldstone boson} $G_{ij}(\vartheta)$ or $W(\vartheta)$.  

Using these choices in eq.\ \pref{PCresult} finally leads to the following inflationary power-counting estimate for a Feynman graph with $\E$ external lines  
\ba \label{PCinf}
 \B_\E (H) &\simeq& \frac{M_p^2}{H^2} \left( \frac{H^2}{M_p} \right)^\E
 \left( \frac{H}{4 \pi \, M_p}
 \right)^{2L} \left\{ \prod_{i,I_s,I_h}  \left[ \prod_{d=0,1,2} \Bigl( \epsilon^{d/2} \; \hat c^{d,i}_{2-d, \ssI_s,\ssI_h} \Bigr)^{V^{d,i}_{2-d,\ssI_s,\ssI_h}} \right] \right\} \\
 &&  \qquad \times
 \left[ \prod_{i,I_s,I_h} \Bigl( \epsilon^{I_s/2} \hat\lambda_{\ssI_s}\Bigr)^{V^{0,i}_{0,\ssI_s,\ssI_h}} \right]
 \prod_{i,I_s,I_h} \left\{ \prod_{d+D \ge 4} \left[ \epsilon^{d/2}\hat g^{d,i}_{\ssD,\ssI} \left( \frac{H}{M_p}
 \right)^2 \left( \frac{H}{M}
 \right)^{d+D-4} \right]^{V^{d,i}_{\ssD,\ssI_s,\ssI_h}} \right\} \,. \nn
\ea
Here the products are over all vertex types appearing in the graph, labeled according to the number of background scalar fields ($i$) and metric ($I_h$) or scalar perturbations ($I_s$) participating in the interaction. The vertices are labelled according to the number of derivatives on background scalar fields ($d$) and the total number of derivatives, $d_n=D+d$, on all background fields or fluctuations. This expression summarizes the $\epsilon$ and $H/M_p$ dependence of a general Feynman graph under simple inflationary assumptions, and so is the main result of this section. 

\subsection{Single-field slow-roll inflation}
\label{ssec:examples}

Before exploring the trade-off between $\epsilon$ and $H/M_p$ in exotic situations, it is worth first verifying that the above rules capture the known dependence of fluctuations in situations already considered in the literature. In making contact with the literature we must become more explicit about the gravitational sector, for which the fluctuation $h_{\mu\nu}$ contains both scalar and tensor parts. At this point we also specialize to single-field models, which in practice means that we can always choose $G_{ij}(\theta)$ to be a constant so that the inflaton kinetic term is proportional to $\sqrt{-g} \; \partial_\mu \phi \,\partial^\mu \phi$, and so does not contain any trilinear or higher inflaton self-interactions (though it does contain trilinear and higher interactions coupling powers of the metric fluctuation to two inflaton fluctuations).

For the purposes of tracking $\epsilon$ it is convenient to work in a gauge where both $\phi$ and the scalar part of $h$ have diagonal $\phi$-$\phi$ and $h$-$h$ kinetic terms unsuppressed by $\epsilon$ while the off-diagonal $\phi$-$h$ kinetic mixing is order $\sqrt\epsilon$. This is precisely the counting one would have in an inflationary model when expanding the inflaton kinetic term $\sqrt{-(\hat g+h)} \; \partial_\mu(\varphi + \phi) \partial^\mu(\varphi + \phi)$ out to quadratic order, using the above counting rules that convert $\dot \varphi \to \sqrt\epsilon \; H M_p$. 

We must also come to grips with the gauge-dependence of the gravitational sector. Since we regard our scalar to be canonically normalized we effectively work in a non-unitary gauge for which the scalar field can be tracked separately from the metric fluctuation, though only one combination of these survives in physical quantities. The result in unitary gauge (and for dimensionless tensor modes, $t_{\mu\nu}$) can be found by the rescaling\footnote{To write the full action at cubic order and higher, we also need the non-linear terms in the gauge transformation \cite{Maldacena}. Including these terms gives, parametrically, $\zeta \simeq  \frac{ \phi}{\sqrt\epsilon \; M_p} (1+\mathcal{O}(\sqrt{\epsilon}\phi/M_p)+\dots)$ and so will not change our leading order results below.}  
\be \label{zetadef}
   \zeta \simeq \frac{ \phi}{\dot \varphi/H} \simeq \frac{ \phi}{\sqrt\epsilon \; M_p} \qquad \hbox{and} \qquad
   t_{\mu\nu} \simeq \frac{h_{\mu\nu}}{M_p} \,.
\ee 

With these rules we expect the leading contribution to the variance of $\phi$ and $h$ to correspond to the lowest-order result for a Feynman graph with $\E = 2$ and $L = 0$ that only uses vertices taken from the 2-derivative interactions. The diagonal terms arise unsuppressed by powers of $H/M_p$ or of $\epsilon$, while as discussed above the off-diagonal terms are down by at least one power of $\sqrt\epsilon$.  The result therefore is of order 
\be
  \B_{hh}(H) \simeq \B_{\phi\phi} \simeq H^2 \,, \quad \hbox{while} \quad \B_{\phi h} \simeq \sqrt\epsilon \; H^2 \,.
\ee  
Converting to curvature fluctuations and dimensionless strain using eq.\ \pref{zetadef} then leads to the usual estimates
\be
   \B_{\zeta\zeta} \simeq \frac{H^2}{\epsilon\, M_p^2}  \qquad \hbox{and} \qquad
    \B_{\ssT\ssT} \simeq \frac{H^2}{M_p^2} \,.
\ee

The leading powers of $H/M_p$ in the bispectra are similarly obtained by choosing $\E=3$ and $L=0$ and no vertices used except those with $d_n =2$.  The leading powers of $\epsilon$ are then found from the estimates using the simplest graph involving only a single 3-point vertex. For the quantities $\langle hhh \rangle$ and $\langle h\phi \phi \rangle$ this leads to the $\epsilon$-unsuppressed estimates
\be\label{eqn:bispecest1}
  \B_{hhh}(H) \simeq \B_{h\phi\phi} \simeq \frac{H^4}{M_p} \,, 
\ee  
since the required unsuppressed cubic vertex comes from either the Einstein-Hilbert action or the inflaton kinetic term. The same is not true for $\langle hh\phi \rangle$ or $\langle \phi \phi \phi \rangle$ since there is no cubic interaction of these types arising unsuppressed by $\epsilon$ in the $d_n \le 2$ lagrangian. Since the cubic scalar interaction is order $\epsilon^{3/2}$ it is subdominant to the interactions obtained by inserting a single $h$-$\phi$ kinetic mixing into $\langle hhh \rangle$ or $\langle hh\phi \rangle$, leading to the estimates 
\be\label{eqn:bispecest2}
  \B_{hh\phi}(H) \simeq \B_{\phi\phi\phi} \simeq \frac{\sqrt\epsilon \; H^4}{M_p} \,.
\ee  
Again converting to dimensionless strain and curvature fluctuation using eq.\  \pref{zetadef} then leads to the usual estimates \cite{Maldacena}
\be
  \B_{\ssT\ssT\ssT}(H) \simeq \B_{\ssT\ssT\zeta}(H) \simeq \frac{H^4}{M_p^4}  \qquad \hbox{and} \qquad
   \B_{\ssT\zeta\zeta}(H) \simeq \B_{\zeta\zeta\zeta}(H) \simeq  \frac{H^4}{\epsilon \, M_p^4} \,.
\ee  
\begin{figure}[t!]
\centering
\includegraphics[width =5 in]{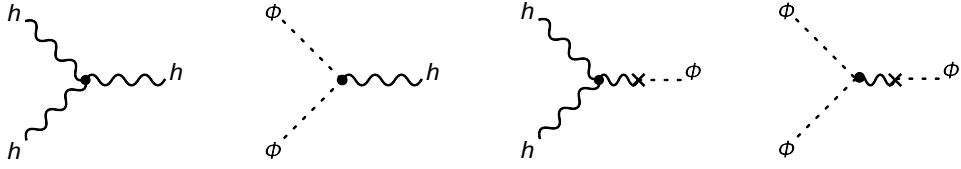}
\caption{Leading Feynman graphs corresponding to the left hand sides of eq.\ \eqref{eqn:bispecest1} and  eq.\ \eqref{eqn:bispecest2}}
\end{figure}
Continuing in this fashion for the tri-spectra, using the tree graphs with $\E = 4$, $L=0$ and $V_n = 0$ unless $d_n = 2$ similarly leads to 
\be
  \B_{hhhh}(H) \simeq \B_{hh\phi\phi} \simeq \frac{H^6}{M_p^2}\quad \hbox{while}\quad 
  \B_{hhh\phi}(H) \simeq \B_{h\phi\phi\phi} \simeq \frac{\sqrt\epsilon \; H^6}{M_p^2} 
  \quad \hbox{and} \quad \B_{\phi\phi\phi\phi} \simeq \frac{\epsilon \, H^6}{M_p^2} \,,
\ee  
and so the leading contributions to the dimensionless strain and curvature perturbation correlations scale as
\be
  \B_{\ssT\ssT\ssT\ssT}(H) \simeq \B_{\ssT\ssT\ssT\zeta}(H) \simeq \frac{H^6}{M_p^6} \quad \hbox{while}\quad \B_{\ssT\ssT\zeta\zeta} \simeq  \B_{\ssT\zeta\zeta\zeta} \simeq \B_{\zeta\zeta\zeta\zeta} \simeq \frac{H^6}{\epsilon \, M_p^6}  \,,
\ee  
and so on to any order, and for any correlation function, desired.

\subsection{Examples near the perturbative boundary}

In this section we turn to several examples for which the previous power-counting points to nontrivial classes of graphs that must be summed to infer reliably the properties of correlations. The purposes of doing so are to highlight the need in these cases for additional arguments in order to ensure the theory retains some predictive power.

\subsubsection{Small sound speed} 
\label{ssec:smallcs}

One example along these lines is when higher-derivative interactions involving the inflaton become important. This limit is normally discussed in terms of a small sound speed, $c_s \ll 1$, since expanding higher-derivative scalar self-interaction using $X = -\Bigl[ \partial(\varphi + \phi)\Bigr]^2 =  \dot\varphi^2 + 2 \dot \varphi \, \dot \phi - (\partial \phi)^2$ produces modifications to the speed of mode propagation, since
\be
 X + \frac{c_{41} \, X^2}{M^4} \supset \left(1 + \frac{2c_{41} \dot\varphi^2}{M^4} \right) \dot\phi^2 - (\nabla \phi)^2 \,,
\ee
and so $c_s^{-2} \simeq 1 + 2c_{41} \dot\varphi^2/M^4$, where $c_{41}$ is a dimensionless effective coupling.

It is clear that in order to obtain $c_s$ much different from unity one must choose $\dot\varphi$ and $M$ to satisfy
\be \label{largephidot}
  \frac{\dot\varphi}{M^2} \simeq \frac{\sqrt\epsilon \; HM_p}{M^2} \simeq \frac{\sqrt\epsilon \; v^2}{M^2} \simeq \O(1) \,,
\ee
and the purpose of this section is to outline the extent to which such a choice poses a threat to the power-counting given above. Clearly a minimal requirement is the validity of the semiclassical expansion itself, but at face value eq.\ \pref{PCinf}  says this requires only $H \ll 4\pi M_p$ and $H \ll M$, neither of which in themselves preclude the condition at eq.\ \pref{largephidot}. 

The effective theory's validity also imposes additional conditions, such as that the background evolution must be adiabatic \cite{EFTnonadiabatic, GREFT, adiabatic}. Since $M$ is the smallest UV scale in the EFT, in the present context adiabaticity implies both $\dot a / a = H \ll M$ (which we already impose) as well as the weaker condition $\dot\varphi/\varphi \simeq \sqrt\epsilon \, H \ll M$. These also seem consistent with \pref{largephidot}.

Does anything preclude the regime in eq.\ \pref{largephidot}? It is clear that to the extent that $\epsilon \ll 1$, eq.\ \pref{largephidot} requires $M \ll v$ (for instance $M \lsim \epsilon^{1/4} v$ would do the job). The issue is whether or not it is legitimate in an EFT to have the universe be dominated by an energy density, $V \simeq v^4$, that is larger than the UV scales being integrated out: $v \gg M$.  

It happens that having an energy density above the UV scale in itself need not rule out the use of EFT methods. It is possible if the large energy density cannot be converted to more dangerous forms that violate the central EFT assumption that the motion involves only the specified low-energy degrees of freedom \cite{UVblock}. Dangerous processes from this point of view are those (for example) that transfer too much kinetic energy to background fields or cause excessive particle production.\footnote{Although examples --- often supersymmetric --- can arise for which a large background energy, $V \simeq v^4$ with $v \gg M$, does not preclude using an EFT whose domain of validity is energies below $M$, these scenarios tend to be special and can break down at later, less protected, points in the history of the universe (such as reheating) \cite{UVblock}.}

But the condition in eq.\ \pref{largephidot} requires more than just that the potential energy $V \simeq v^4 \gg M^4$, it also demands the scalar {\it kinetic} energy be of order (or larger than) UV scales, since $\dot\varphi^2 \gsim M^4$. What makes this precarious is that this kinetic energy must not be extractable to excite either UV particles (by assumption, not in the EFT) or to provoke non-adiabatic background evolution (which again is not reliably captured by the EFT). 

Were it not for an explicit example one would be tempted to conclude from all this that small sound speeds must be beyond the domain of sensible EFT methods. 
 
\subsubsection*{DBI inflation}

The explicit example that seems to argue otherwise is DBI inflation \cite{DBI}, for which the inflaton arises as the centre-of-mass coordinate of a brane, for which it is argued that relativistic kinematics implies an action with a kinetic term of the form
\be \label{DBI}
 \L = - \sqrt{-g}\; T \sqrt{1 - X/T} \,,
\ee
where $X = -(\partial \phi)^2$ and $T$ is the brane's tension. 

What is unusual about the DBI action is that it keeps all orders in $X$ but neglects all second and higher derivatives of $\phi$, and this is believed to be a sensible regime because relativistic kinematics should not break down even for speeds near the speed of light --- which in this case corresponds to the limit $|X/T| \simeq \O(1)$. At the classical level this is self-consistent, inasmuch as the classical equations coming from eq.\ \pref{DBI} drive $\ddot\varphi \to 0$ as $\dot\varphi^2 \to T$. And in the regime $|X/T|\simeq \O(1)$ the speed of sound predicted for the inflaton becomes small, as expected from the above estimates.

The special structure of the DBI action has been argued to be preserved by a symmetry (nonlinear realization of the spacetime symmetries that are broken by the presence of the brane) \cite{DBIsym, DBIsym2}, and this symmetry is likely to ensure the presence of the special square-root factor in front of all effective interactions, including those involving higher derivatives of $\phi$. It is quite possible that this, together with the use of a different scaling regime for which spatial derivatives are of order $H/c_s$, might lead to a consistent power-counting formulation\footnote{Notice that the power-counting issue differs from the issue of the technical naturalness of small $c_s$ discussed in ref.\ \cite{TNSmallcs}. Technical naturalness asks whether having small $c_s$ at one scale ensures it remains small when run to other scales, for which the dangerous interactions are usually those involving the heaviest fields (which by assumption are not present in the EFT). By contrast, power-counting questions whether or not having small $c_s$ undermines the entire low-energy expansion on which semiclassical methods are based.} by extending the preliminary steps taken in ref.\ \cite{SmallcsPC} (see also \cite{SmallcsBounce}).

Our discussion above shows how worthwhile it would be to develop such a systematic power-counting calculation (along the lines of the above) for DBI (and, by extension, some class of small-$c_s$ models), showing how semiclassical calculations can be regarded as the leading terms in the expansion in powers of a small parameter (and, if so, explicitly what is this parameter). In the absence of such a power-counting result it is difficult to know how to quantify precisely the theoretical error made when using such semiclassical methods, and thereby to know how far to trust its predictions.

Although the existence of such a power-counting scheme would mean that a DBI-type action could be self-consistent, it would not automatically guarantee that it provides a good description for any specific microscopic brane setup. This is because there are usually additional issues that need checking, as can be seen by keeping in mind the simple example of a first-quantized relativistic point particle. The Nambu-like action for such a particle is very similar to the DBI action (due to the similarly broken spacetime symmetries), and would at first sight seem equally tricky to power-count when in the extreme relativistic limit (for which the centre-of-mass coordinate, $\bfr(t)$, satisfies $\sqrt{1 - \dot \bfr^2} \to 0$, similar to the relativistic DBI limit $\sqrt{1-X/T} \to 0$). Yet we know that a consistent power-counting formulation in this case exists, and is most easily given in the second-quantized framework obtained once the relevant antiparticle is also included. In order for an EFT description cast purely in terms of the centre-of-mass motion to be valid one must check whether or not the relativistic motion starts to produce particle-antiparticle pairs, or to radiate other states to which the moving particle couples. Similarly, for relativistic DBI constructions one must identify the extent to which any assumed relativistic motion similarly stimulates brane, string or Kaluza-Klein excitations that have been assumed to be integrated out when formulating the EFT involving only $\phi$ and its derivatives, but neglecting these other states \cite{CheckOtherModes}.

\subsubsection{Eternal inflation}
\label{ssec:eternal}

Our power-counting discussion shows that classical dynamics always dominates when $H/M_p$ is sufficiently small, regardless of the size of $\epsilon$. Working classically to subdominant order in $\epsilon$ therefore implicitly assumes a hierarchy of the form $H/M_p \ll \epsilon^s$ for some positive $s$, whose value depends on precisely how far one wishes to work in the slow-roll expansion. 

But for any specific value of $\epsilon$ and $H/M_p$ it is also clear that beyond some order in $\epsilon$ it becomes invalid to work purely at classical order. For instance, reasonable values for $H$ and $\epsilon$ might be $H/(\sqrt\epsilon\,M_p) \simeq 10^{-5}$ and $\epsilon \simeq 10^{-2}$, for which $H/M_p \simeq \epsilon^{3}$. In this case working beyond 6th order in $\epsilon$ necessarily also requires including loop corrections and including the action's higher-derivative terms in any classical calculation.

An extreme example to which this observation is pertinent is the case of eternal inflation, which corresponds to choosing parameters so that 
\be \label{eternal}
  \frac{\delta \phi }{ \dot\varphi/H} \simeq \frac{H}{4 \pi \sqrt\epsilon \, M_p} \gsim \O(1) \,.
\ee
This condition ensures that inflationary stochastic fluctuations can compete with classical evolution over Hubble time-scales. Clearly, because $\epsilon \lsim (H/4\pi M_p)^2$ in this regime contributions suppressed by $\epsilon$ at any loop order (say, tree level) can compete with contributions unsuppressed by $\epsilon$ but at one higher loop.

As has been pointed out elsewhere, the eternal-inflation regime is parametrically consistent with a perturbative analysis (although not if $c_s$ is too small \cite{eternalptbv, SmallcsPC}) since both control parameters $H/M_p$ and $\epsilon$ can be arbitrarily small while still satisfying eq.\ \pref{eternal}. Where the above power-counting becomes interesting is if effects are computed for which $\epsilon$ being nonzero plays a role, implying the keeping of a fixed order in the $\epsilon$ expansion. Power-counting then shows that once one keeps {\it any} terms linear in $\epsilon$ (or smaller), it becomes inconsistent to work only within the classical approximation, using only up-to-two-derivative interactions, such as those of General Relativity coupled to an inflaton.

For stochastic formulations of eternal inflation \cite{StochInf} these arguments indicate that inclusion of drift is only consistent, while neglecting quantum and higher-derivative corrections to General Relativity, in the regime
\be
   \frac{H}{4\pi M_p}  \gsim \sqrt\epsilon \gsim \left( \frac{H}{4\pi M_p} \right)^2 \,,
\ee
where the first inequality restates the condition in eq.\ \pref{eternal} for eternal inflation and the second inequality is the condition that the drift --- normally proportional to $V' \propto \sqrt\epsilon$ --- be larger than one-loop corrections. It also means that all corrections to the noise (and subdominant contributions to the drift) arising at $\O(\epsilon)$, such as those found in ref.\ \cite{StochCorr}, must also be accompanied by loop and higher-derivative corrections if applied within the eternal-inflation regime. 

In view of recent resurgence of interest in stochastic methods \cite{StochCorr,StochDecoh,RecentStoch} it clearly would be worthwhile developing systematic power-counting tools of equal power for the stochastic regime.

\section*{Acknowledgements}
We thank Subodh Patil, Eva Silverstein, Andrew Tolley, and David Tong for helpful discussions about power-counting and small-$c_s$ models. We also thank the Banff International Research Station for support and hospitality while this work was in progress. 
This work was partially supported by funds from the Natural Sciences and Engineering Research Council (NSERC) of Canada. Research at the Perimeter Institute is supported in part by the Government of Canada through NSERC and by the Province of Ontario through MRI. The work of PA was supported in part by the United States Department of Energy through grant DE-SC0015655.

\end{document}